\documentclass{aa}  
\usepackage{graphicx}
\usepackage{txfonts}

\usepackage{ifthen}
\newboolean{ARXIVVERSION}
\setboolean{ARXIVVERSION}{true}

\def\PC{{\rm pc}}
\def\KPC{{\rm kpc}}
\def\KMS{{\rm km\,s}^{\rm -1}}

\def\MSUN{{\rm M}_\odot}

\def\ERR#1{^{\pm #1}}	

\def\THINGS#1#2#3#4{%
\includegraphics[width=5.2cm]{#1#2.eps} &
\includegraphics[width=5.2cm]{#1#3.eps} &
\includegraphics[width=5.2cm]{#1#4.eps}
}
\def\FIGS#1#2#3#4{%
\includegraphics[width=#1cm]{#2} &
\includegraphics[width=#1cm]{#3} &
\includegraphics[width=#1cm]{#4}
}

\begin{document}

\title{The Bosma effect revisited}
\subtitle{I. HI and stellar disc scaling models}

\author{
	Frederic V. Hessman
	\and
	Monika Ziebart
	}


\institute{
	Institut f\"ur Astrophysik, Georg-August-Universit\"at, Friedrich-Hund-Platz 1, D-37077 G\"ottingen
             }

\date{Received 6 May 2011; accepted 17 June 2011}
\abstract
{
The observed proportionality between the centripetal contribution of the dynamically insignificant HI gas in the discs of spiral galaxies and the dominant contribution of Dark Matter (DM) -- the ``Bosma effect'' -- has been repeatedly mentioned in the literature but largely ignored.
Since this phenomenology, if statistically significant, tells us something about the relationship between the visible baryonic and invisible DM, it is important to re-examine the reality of this effect using formal tests and more modern data.
}
{
We have re-examined the evidence for the Bosma effect, either by scaling the contribution of the HI gas alone or by using both the observed stellar disc and HI gas as proxies.
}
{
We have calculated Bosma effect models for 17 galaxies in The HI Nearby Galaxy Survey data set.
The results are compared with two models for exotic cold DM: internally consistent cosmological Navarro-Frenk-White (NFW) models with constrained compactness parameters, and  "Universal Rotation Curve" (URC) models using fully unconstrained Burkert density profiles.
}
{
Fits to spiral galaxy rotation curves computed using just HI scaling are inadequate, despite the clear proportionality seen in the outer discs.
The poor performance is obviously related to the prominent decrease in the HI surface density in regions of high stellar surface density, where HI gas been converted into molecules and stars.
The Bosma models that partially correct for this physical effect using the stellar discs as additional proxies are statistically nearly as good as the URC models and clearly better than the NFW ones.
}
{
We confirm the correlation between the centripetal effects of DM and that of the interstellar medium of spiral galaxies.
The edificacy of ``maximal disc'' models is explained as the natural consequence of ``classic'' Bosma models which include the stellar disc as a proxy in regions of reduced atomic gas.
The perception that the Bosma effect could be due to the near-equality of the HI surface density and the projected mass density of a cold DM halo is incorrect, both theoretically and empirically.
The standard explanation -- that the effect reflects a statistical correlation between the visible and exotic DM -- seems highly unlikely, 
given that the geometric forms and hence centripetal signatures of spherical halo and disc components are so different.
A literal interpretation of the Bosma effect as being due to the presence of significant amounts of disc DM requires a median visible baryon to disc DM ratio of about 40\%.
}

\maketitle
%

\section{Introduction}

One of the pillars supporting the present paradigm of cold Dark Matter (hereafter CDM) is the interpretation of the rotation curves of spiral galaxies: the latter are thought to require the presence of radially extended, quasi-spherical and very massive halos of exotic non-baryonic matter
(Sanders \cite{978-0521113014}).
Nevertheless, there still remain many unanswered questions and inconsistencies between theory and observations, particularly about the role and effects of the baryons.
While the signs of Dark Matter (hereafter DM) are particularly evident in massive galaxies with nearly flat rather than Keplerian outer rotation curves (assuming that the mass-to-light ratio of the disc is not an increasing function of radius), the most extreme effect of DM is actually seen in dwarf spirals with slowly rising rotation curves and minimal stellar and small gaseous contributions (Moore \cite{1994Natur.370..629M}).
Fits to the rotation curves of both types of galaxies using the observed stellar and gaseous components and various models for the CDM halo have shown that the data are not well-fit using the theoretically expected Navarro-Frenk-White (Navarro, Frenk \& White \cite{1997ApJ...490..493N}; hereafter NFW) profiles, whose central density profiles are too ``cuspy'' (de\,Blok \& McGaugh \cite{1997MNRAS.290..533D}) and simple adiabatic contraction of the CDM with the baryons should result in even more cuspy profiles (Sellwood \& McGaugh \cite{2005ApJ...634...70S}).
Given enough interaction between the baryons and CDM halos during the formation phase of galaxies, it is claimed that this difference can be explained (Navarro, Frenk \& White \cite{1996ApJ...462..563N}; Mashchenko, Couchman \& Wadsley \cite{2006Natur.442..539M}; El-Zant, Shlosman \& Hoffman \cite{2001ApJ...560..636E}; Tonini, Lapi \& Salucci \cite{2006ApJ...649..591T}; Governato et al. \cite{2010Natur.463..203G}),
but there is still a particular problem for dwarf galaxies, since the minimal amount of baryonic mass is unlikely to have been able to affect the dynamically dominant CDM halo, so still {\it ad hoc} models for the dissipative effects of baryons have to be invoked (Oh et al. \cite{2010arXiv1011.2777O}).

There are other signs of deep and empirically simple but theoretically unexpected and largely unexplained connections between the visible baryonic and invisible dark components: 
the {\it MOdified Newtonian Dynamics} phenomenology (MOND; Milgrom \cite{1983ApJ...270..365M}; Sanders \& McGaugh \cite{2002ARA&A..40..263S}; Gentile, Famaey \& de\,Blok \cite{2011A&A...527A..76G}); 
the mass-discrepancy relation (McGaugh \cite{2004ApJ...609..652M});
the baryonic Tully-Fischer relation (McGaugh \cite{2005ApJ...632..859M});  and
the universal mean surface densities derived both for CDM halos (Gentile et al. \cite{2009Natur.461..627G}) and baryonic discs (Donato et al. \cite{2009MNRAS.397.1169D}).

The most unusual correlation between the observed properties of disc galaxies and the derived properties of the dark components was identified by Bosma (\cite{1978PhDT.......195B}, \cite{1981AJ.....86.1825B}): the centripetal contribution to the global rotation curves by the dynamically insignificant gaseous discs appears to be directly correlated with that of the remaining DM component -- the ``Bosma effect''.
This effect has been characterized in various ways by different authors.
After deriving the total disc surface densities implied by the rotation curves, $\Sigma_{dyn}(R)$, and comparing them with the observed HI or HI+H$_2$ gas surface densities, $\Sigma_{gas}(R)$, Bosma observed ``{\it ... that as a rule the ratio} [of the two] {\it ... is more or less constant beyond about one-third of the optical radius, with} [HI] {\it being the dominant contributor ... in the outer parts}'' (Bosma  \cite{1981AJ.....86.1825B}).
Sancisi (\cite{1999ApSS.269...59S}) simplified the effect by stating that ``{\it ... the rotation curve in the outer parts can be explained ... by scaling up the HI disc by about a factor} [of] {\it 10}''.
Hoekstra, van\,Albada \& Sancisi (\cite{2001MNRAS.323..453H}; hereafter HvAS) studied the Bosma effect in a sample of normal galaxies: they said that ``{\it ... the total surface density of matter needed to explain the observed rotation curves ...} [is] {\it roughly proportional to the surface density of neutral hydrogen}'' and proceeded to define the proportionality as
\begin{equation}
\label{equation:fHI}
V^2_{DM}(R) ~ \approx ~ \left(\frac{\Sigma_{DM}}{\Sigma_{HI}} \right) V^2_{HI}(R)
\end{equation}
HvAS and the others explicitly defined the constant of proportionality between the centripetal effects of the HI gas and the DM as the assumed constant ratio of the DM to HI surface densities averaged over the same disc, corrected only for the presence of Helium.
Bosma, however, originally explicitly meant to include the total gas surface density, even when he used HI as a proxy for the interstellar medium (ISM), so there are potentially several different versions of this effect, depending upon what assumptions are made:
\begin{itemize}
\item what we will call the ``simple'' Bosma effect or ``pure HI-scaling'' uses only $\Sigma_{HI}$, implicitly or explicitly corrected for the contribution of He and heavy elements; whereas
\item the ``classic'' Bosma effect attempts to include the total gaseous surface density, either explicitly by using $\Sigma_{ISM}\!=\!\Sigma_{HI}+\Sigma_{H_2} $ (again corrected for He and heavy elements), or by using the stellar disc as an additional proxy.
The latter can be done by allowing the effective stellar mass-to-light ratio, $\Upsilon_{disc}$, to be a free parameter.
\end{itemize}

The mean HI Bosma ratio has been measured in almost a hundred normal and dwarf spiral galaxies spanning a wide range of Hubble types from massive early spirals to late dwarf irregulars (Bosma \cite{1981AJ.....86.1825B}; Carignan \cite{1985ApJ...299...59C}; Carignan \& Beaulieu \cite{1989ApJ...347..760C}; Carignan \& Puche \cite{1990AJ....100..641C}; Jobin \& Carignan \cite{1990AJ....100..648J}; Puche, Carignan \& Bosma \cite{1990AJ....100.1468P}; Swaters \cite{1999PhDT........27S}; HvAS; Noordermeer \cite{2006PhDT.........1N}).
A direct correlation between the gravitational contributions of the gas and DM is seen in practically all galaxies analyzed (Sancisi \cite{1999ApSS.269...59S}; Bosma \cite{2004IAUS..220...39B}).
While Swaters (\cite{1999PhDT........27S}) found value of $3-5$ (corrected for He) in ``late-type'' dwarf spiral galaxies, HvAS found a wide range of $5\!-\!20$ in normal galaxies with a median value of about 7 and Noordermeer (\cite{2006PhDT.........1N}) a range from $5\!-\!52$ and a median value of 17 in early spirals.
This behavior is supported by a correlation between the value of the Bosma factors and the masses of galaxies found by Hoektra et al.: more massive galaxies tended to have larger values.
With just a few exceptions, the rotation curves obtained by scaling up the centripetal contribution of the HI gas by some constant factor and not including a spherical DM halo resulted in model rotation curves which followed the observed curves about as accurately as a CDM halo model permits.

HvAS explicitly set out to test the reality of the Bosma effect.
After admitting that ``{\it ...most of the 24 rotation curves can be fitted rather well over their full extent}'', they then proceeded to explain it away using various arguments, so that the effect is now largely ignored in the literature.
A critical review of the arguments suggests that their conclusion was premature:\\[-3ex]
\begin{itemize}
\item ``{\it The model curve} [of the poorer fits] {\it does not agree with the observed rotation curve in the inner region.}''
However, they admitted that their treatment of the central bulge, beam smearing, and other errors could be the cause of this problem.
%
\item There are ``{\it ... large wiggles that are not present in the observed rotation curve.}''
While this is true in a few cases, they admit that the wiggles can be created by spiral arm structures in which the HI can be converted to H$_2$ and other molecules.  The HI surface density may, in fact, be a very ``noisy'' tracer of the global gas surface density in galaxies.  This is easily illustrated in the ``holes'' seen in spatially well-resolved HI images of face-on galaxies (e.g. Deul \& Den Hartog \cite{1990AA...229..362D}).
In fact, many rotation curves show small-scale structure which cannot easily be interpreted as being due to mass-density variations using any model.
\item ``{\it The model rotation curve drops below the observed rotation curve at large radii.}''
The outer scaled-up HI rotation curves ended up significantly ($\ga 10\%$) below the observed ones in about 7-8 of the 24 galaxies, but the deviation always occurs at the very outer edge of the visible HI disc ($93\!\pm\!9$\%).
HvAS admitted that there can be several reasons for this effect (e.g. large scale flux not seen by the interferometers, partial ionization), leading them to conclude that ``{\it ... the observed HI surface density may therefore not always trace the total gas density in the outer parts}.''
Additionally, the HI rotation curves are almost always calculated from the observe surface densities assuming that the disc is effectively infinite,
systematically minimizing the rotation curve velocity at the edge of the discs (Casertano \cite{1983MNRAS.203..735C}).
\item ``{\it ... scaling of HI to represent the dark component only works in combination with maximal discs}'' (Palunas \& Williams \cite{2000AJ....120.2884P}).
This effect is not surprising, since the models using HI alone could simply be correcting for the missing H$_2$ gas component most strongly associated with the stars.
\item ``{\it ... our sample is biased against galaxies with $R_{out}/h_{HI}$ substantially larger than 3.}''  This argument is based on the assumption that a large disc exponential scalelength is needed to insure the flatness of the rotation curves at large radii.   However, the observed HI distributions in the HvAS sample are not particularly exponential: the observe distributions are better described as being flat in the inner parts, with nearly constant surface densities of $5-10\,\MSUN/\PC^2$ and having complex tails (to what extent the total gas surface densities are exponential is unknown).
They suggested that better observations would eventually show the drop in the scaled HI curve and hence the failure of the Bosma effect, but Swaters (\cite{1999PhDT........27S}) found no such problem in galaxies where the extents of exponential discs are clearly much larger than those of the stellar discs.
\item Although HvAS admit that ``{\it .. for about two-thirds of the galaxies we obtain good fits to the data}'', one has the impression that in fact practically all of the galaxies are well fit by the Bosma effect.
Since the HvAS fits were done by eye and were not explicitly compared with standard CDM rotation curve fits, it is impossible to judge how bad or good a ``2/3'' success rate really is.
\item They finally argue against the reality of the Bosma effect fits,  saying that ``{\it the good fits are somewhat coincidental}'' because they expect the HI curves to eventually decline very rapidly, even though their own data did not, in fact, show this.
\end{itemize}

HvAS finally conclude that ``{\it ... in many cases,, there seems to be little or no relation between HI and dark matter}'' and ''{\it ... it is not possible to conclude here that there is a real coupling between HI and dark matter in spiral galaxies.}''
These negative conclusions are obviously very conservative: measured on the standards of typical rotation curve fits using any mass model, the results were, in fact, amazing good.
Given the success of the phenomenological model -- there is no way to modify the shape of the centripetal contribution except by global scaling of the stars and HI -- and a natural explanation for the cases that didn't fit quite as well, HvAS could in fact not show the Bosma effect is a statistical fluke or selection effect and hence to disprove that there was {\em no} coupling between the ISM and the gravitational effects attributed to DM in spiral galaxies.

\begin{figure}
\resizebox{\hsize}{!}{\includegraphics{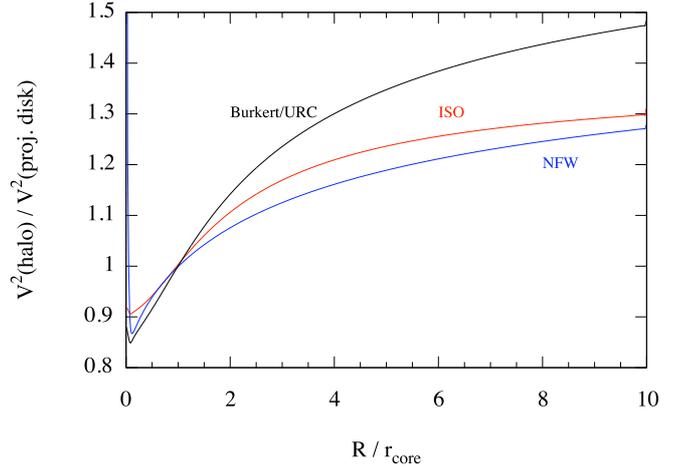}}
\caption{A comparison of the centripetal effects of spherical halos with compactness $c=10$ and the corresponding discs with the same vertically projected surface mass-densities.  The ratios are arbitrarily normalized to the values at $R=r_{core}$ to emphasize the similarities in shape.}
\label{figure:halodisk}
\end{figure}

At the time Bosma first described this correlation, placing the DM in a disc was a reasonable assumption, but already at the time of HvAS, the DM was assumed to reside in a quasi-spherical halo.
Since the centripetal acceleration $V^2/R$ at a given radius is calculated only indirectly from an integral over the {\em entire} surface mass density for discs and just over the {\em inner} volume mass density for spherical halos, the connection between the surface or volume densities and the rotation curve contributions is dramatically different for the two geometries -- unlike the spherical case, the gravitational effects of matter distributed in a disc-like manner effects the motions of test particles both within and without the disc, making it much more difficult to obtain a particular rotation curve contribution.
Ignoring this distinction has led to considerable confusion in the literature about the role of surface density: e.g. Meuerer \& Zheng (\cite{2011AAS...21724613M}) incorrectly characterize the Bosma effect as showing that ``{\it ... the HI column density in extended HI discs often is nearly a linear tracer of the projected DM surface density}.''
The centripetal effect of a spherical halo is, in general, not at all the same as that of a disc having a surface density equal to the projected density of the spherical halo.
The normalized ratios of the centripetal effects of three different spherical halos and their equivalent projected discs are shown in Fig.\,\ref{figure:halodisk}:
while there is, of course, some similarity in shape, the relative differences for the regions relevant for the HI discs are about 30\%, with gradients in the relative effects systematically largest in the inner regions most relevant to the Bosma effect.
The only exception to this rule is that the asymptotic projected surface density and $V(R)$ of a singular isothermal halo is the same as that of an infinite disc with $\Sigma(R)\propto R^{-1}$.
However non-singular isothermal CDM halo fits yield small but significant core radii (this is one of the main arguments against ``cuspy'' NFW profiles; de\,Blok et al. \cite{2008AJ....136.2648D}), HI discs are not infinite and they do not in general show this surface density scaling (see Fig.\,\ref{figure:sigmas}).
Thus, the Bosma effect has nothing to do with projected CDM density profiles.

In summary, the usual characterisation of the Bosma effect in the literature is wrong on three counts:
\begin{itemize}
\item 
the implied physical correlation is between the DM and the total ISM, not just its neutral part; the HI has simply been used as the most directly observable tracer, despite its obvious drawbacks, so that the quality of the observed effect might be much better (or worse) than previously claimed;
\item
the actual comparison is of the centripetal acceleration contributions of gas and DM at each radius, not the surface densities and certainly not the projected surface densities of spherical halos; and
\item
within the standard CDM paradigm, the implied correlation is between two mass components with different geometries, with different projected surface densities, and different centripetal patterns.
\end{itemize}
With this correct description, the Bosma Effekt must be defined phenomenologically as
\begin{equation}
\label{equation:fISM}
V^2_{DM}(R) / V^2_{gas}(R) ~ \approx {\it const}
\end{equation}
where the constant can be interpreted as an effective mass-to-light ratio contribution above that normally attributable to standard ISM components
and the origin of the effect is {\em a priori} not at all obvious.
Unfortunately, previous studies have not shown how the Bosma effect performs relative to standard CDM models.

If real, the Bosma effect obviously has something to say about the relationship between the ISM and DM in disc galaxies.
Fortunately, we can now probe the Bosma effect with much better means: the quality of the HI rotation curves and surface density maps has increased; we can model the stellar populations more accurate due our ability to constrain the stellar contributions in wavelength bands suffering less from interstellar absorption; and H$_2$ maps can be constructed from CO observations with millimeter interferometers.

This article is the first in a series looking again at the Bosma effect.
In this first paper, we compare the performance of ``simple'' (HI-scaling alone) and ``classic'' (stellar disc and HI-scaling) Bosma effect models with those of standard CDM using a set of particularly well-observed galaxies in {\em The HI Nearby Galaxy Survey} (Walter et al. \cite{2008AJ....136.2563W}; hereafter THINGS).

\section{Rotation curve models}

We fit the data with a standard rotation curve model consisting of a stellar disc, a stellar bulge, HI disc, and CDM halo, each baryonic component being assigned a mass-to-light ratio $\Upsilon$:
\begin{equation}
\label{equation:model}
V^2_{obs}  =  \Upsilon_{disc} V^2_{disc}  + \Upsilon_{bulge} V^2_{bulge} +  \Upsilon_{HI} V^2_{HI} + V^2_{CDM}
\end{equation}

The centripetal effects of the ``simple'' Bosma effect -- pure HI-scaling -- are best described by attributing the observed atomic hydrogen with an effective mass-to-light ratio
\begin{equation}
\label{equation:HI}
\Upsilon_{HI} \equiv 1.39 +\Delta\Upsilon_{HI} \equiv 1.39 (1 +f_{HI})
\end{equation}
where the factor of $1.39$ corrects the visible gaseous component for the presence of He and heavier elements (given the roughness of the Bosma model and the uncertainties and systematic errors in the HI data, there is no point in worrying about whether a correction to primordial or to Solar heavy element abundances is more appropriate).
Thus, the Bosma contribution to the effective mass-to-light ratio, $\Delta\Upsilon_{HI}$, can be more usefully defined relative to the total corresponding mass density of the neutal gaseous medium with the scaling factor $f_{HI}\equiv\Delta\Upsilon_{HI}/1.39$.
This is the standard definition for the Bosma effect.

For the ``classic'' Bosma effect fits, the stellar disc is used as a proxy for those inner regions where the density of HI drops far below that which one would have extrapolated from outside, resulting in an effective stellar disc mass-to-light ratio
\begin{equation}
\label{equation:disk}
\Upsilon_{disc} \equiv \Upsilon_{\star}+\Delta\Upsilon_{disc} \equiv \Upsilon_{\star}(1+f_{disc})
\end{equation}
The first term is the correction for the true stellar mass-to-light ratio and the second, again, is the effect due to using the stars as proxies for other -- presumedly non-stellar -- mass components.
The reality of $\Delta\Upsilon_{disc}$ obviously depends critically on the reliability of $\Upsilon_{\star}$.
In the case that $\Upsilon_{\star}$ is derived from infrared images using detailed stellar population models calibrated by observed infrared colours, the true stellar contribution should be fairly reliable.
Again, the ``classic'' Bosma effect measured with the stellar disc as a proxy is best expressed as a scaling of the total stellar mass in the disc, $f_{disc}\equiv\Delta\Upsilon_{disc}/\Upsilon_{\star}$.

In order to quantify the relative success of the Bosma effect at explaining rotation curves of disc galaxies with respect to normal CDM halo models, we have also fit our data with two plausible CDM models: 1) internally consistent theoretical NFW halos and 2) the generic Burkert density profiles (Burkert \cite{1995ApJ...447L..25B}) used by ``Universal Rotation Curve'' assumption (Salucci et al. \cite{2007MNRAS.378...41S}; hereafter URC).

The centripetal contribution of a CDM halo with a NFW density profile is
\begin{equation}
V_{NFW}^2(R)  =  \frac{V^2_{200}}{X} \left[ \frac{ \ln(1+c X)-\frac{c X}{1+c X} }{ \ln(1+c)-\frac{c}{1+c} } \right]
\end{equation}
where $V_{200}$ is the rotational velocity at the so-called ``virial radius'', $r_{200}$, where the mean density of the halo reaches a value 200 times that of the mean cosmic mass-density, the concentration factor $c \equiv r_{200}/r_c$ is the ratio of $r_{200}$ to the NFW core scalelength $r_c$, and where $X\!\equiv\!R/r_{200}$.
Normally, cuspy NFW profiles are characterized by 2 model parameters; this is the approach taken by de\,Blok et al. (\cite{2008AJ....136.2648D}).
However, the N-body calculations which resulted in the NFW profile model clearly show that $c$ is not an independent parameter but is in fact strongly correlated with $V_{200}$ (or, equivalently, with $M_{200}$ or $r_{200}$):
\begin{equation}
c_{NFW} \approx 7.80 \left( \frac{V_{200}}{100\,\KMS} \right)^{-0.294}
\end{equation}
(Maccio, Dutton \& van\,den\,Bosch \cite{2008MNRAS.391.1940M}; their Eqn.\,10).
Including this intrinsic correlation reduces the number of fit parameters by one.
This approach automatically removes the problems with the NFW fits encountered by de\,Blok et al, who found unphysical values of $c$ in 9 of their 20 NFW fits.

The URC model uses a Burkert density profile, resulting in
\begin{equation}
V^2_{URC}(R) =  \frac{V^2_{200}}{X}  \left[ \frac{\ln(1\!+\!(cX)^{2})\!+\!2\ln(1\!+\!cX)\!-\!2\tan^{-1}(cX)}{\ln(1\!+\!c^2)\!+\!2\ln(1\!+\!c)\!-\!2\tan^{-1}(c)} \right]
\end{equation}
In contrast with the NFW model, URC and isothermal 
halo models are usually characterized by the central mass-density $\rho(0)$ and the radial scale length of the core, $r_c$.
This distinction is unnecessary and even misleading, since the virial radii and concentration factors are equally well-defined, making a comparison between the models and the fitted parameters more difficult.
Thus, we parameterize both CDM models with the same quantities, choosing the most commonly used variables $V_{200}$ and $c$.

A summary of the model parameters for each of the four models is given in Table\,\ref{table:parameters}.
Note that each model has very different constraints: the Bosma models have shapes that are independent of the velocity amplitude (other than the possibilities of linear combinations in the ``classic'' version); the shape of the the NFW model is determined by its limiting amplitude; and the URC model has no constraints beyond the scaleable generic shape.
Thus, one naively expects the Bosma models to perform worst and the URC to perform best.

\begin{table}
\caption{Model parameters}
\label{table:parameters}
\centering
\begin{tabular}{c c c c}
\hline\hline
Model             & $f_{disc}$  & $f_{HI}$     & CDM parameter(s) \\
\hline
``Simple'' Bosma  & $\equiv 0$  & $> 0$        & N/A \\
``Classic'' Bosma & $> 0$       & $> 0$        & N/A \\
Constrained NFW   & $\equiv 0$  & $\equiv 0$   & $V_{200}$ \\
Burkert (URC)     & $\equiv 0$  & $\equiv 0$   & $V_{200},c$ \\
\hline
\end{tabular}
\end{table}


\begin{table*}
\caption{The THINGS Sample}
\label{table:sample}
\centering
\begin{tabular}{l l r | r r r | r r r r | r r}
\hline\hline
Name      & Type       & D (Mpc) & $M_\star$ & $M_{bulge}$ & $M_{HI+He}$ & $R_{outer}$ & $h_{HI}$  & for $R>$  & $R_{outer}/h_{HI}$ & $R_{25}$ & $h_{disc}$ \\
\hline
\object{DDO154}  & IB(s)m     &  4.3    & 0.0019    &             & 0.044       & 8.3         & 2.3       & 2.0       & 3.6                & 1.9      & 1.1 \\
\object{IC2574}  & SAB(s)m    &  4.0    & 0.074     &             & 0.81        & 11.7        & 1.2       & 7.7       & 9.8                & 7.7      & 2.9 \\
\object{NGC925}  & SAB(s)d    &  9.2    & 0.72      &             & 0.44        & 13.1        & \multicolumn{3}{c|}{(totally flat)}         & 14.0     & 3.9 \\
\object{NGC2366} & IB(s)m     &  3.4    & 0.018     &             & 0.081       & 8.2         & 1.5       & 4.1       & 5.5                & 4.0      & 2.8 \\
\object{NGC2403} & SAB(s)cd   &  3.2    & 0.33      & 0.029       & 0.36        & 18.0        & 4.9       & 7.7       & 3.7                & 10.2     & 1.7 \\
\object{NGC2841} & SA(r)b     & 14.1    & 7.6       & 1.8         & 1.9         & 51.7        & 15.1      & 23.0      & 3.4                & 16.7     & 3.8 \\
\object{NGC2903} & SAB(rs)bc  & 8.9     & 1.0       & 0.15        & 0.89        & 29.4        & 5.8       & 14.0      & 5.1                & 16.3     & 2.5 \\
\object{NGC2976} & SAc pec    & 3.6     & 0.13      &             & 0.012       & 2.54        & 0.7       & 1.8       & 3.6                & 3.1      & 0.7 \\
\object{NGC3031} & SA(s)ab    & 3.6     & 4.9       & 0.91        & 0.41        & 14.8        & 6.3       & 12.2      & 2.3                & 14.1     & 2.7 \\
\object{NGC3198} & SB(rs)c    & 13.8    & 2.0       & 0.20        & 2.1         & 37.8        & 15.0      & 10.3      & 2.5                & 17.1     & 3.1 \\
\object{NGC3521} & SAB(rs)bc  & 10.7    & 8.7       &             & 1.3         & 31.2        & 6.1       & 17.0      & 5.0                & 17.1     & 4.9 \\
\object{NGC3621} & SA(s)d     & 6.6     & 1.4       &             & 1.3         & 25.8        & 3.7       & 17.0      & 6.8                & 11.8     & 2.5 \\
\object{NGC4736} & (R)SA(r)ab & 4.7     & 1.3       & 0.36        & 0.53        & 9.6         & 4.8       & 2.0       & 2.0                & 7.7      & 1.9 \\
\object{NGC5055} & SA(rs)bc   & 10.1    & 8.7       & 0.65        & 1.4         & 48.8        & 23.8      & 13.0      & 2.1                & 18.5     & 3.9 \\
\object{NGC6946} & SAB(rs)cd  & 5.9     & 4.2       & 0.27        & 0.54        & 19.2        & 7.4       & 9.3       & 2.6                & 9.9      & 3.1 \\
\object{NGC7331} & SA(s)b     & 14.7    & 12        & 1.2         & 1.6         & 25.3        & 3.4       & 20.0      & 7.2                & 22.4     & 4.5 \\
\object{NGC7793} & SA(s)d     & 3.9     & 0.19      &             & 1.2         & 7.7         & 1.7       & 5.0       & 4.5                & 5.3      & 1.5 \\
\hline
\end{tabular}\\
\tablefoot{Masses in units of $10^{10}\,\MSUN$, radii and scalelengths in kpc}
\end{table*}

\section{The sample}

We use the same subset of THINGS galaxies chosen by de\,Blok et al. (\cite{2008AJ....136.2648D}) for their detailed mass models.
Although Gentile, Famaey \& de\,Blok (\cite{2011A&A...527A..76G}) studied an even smaller subset of the THINGS galaxies -- those suffering least from non-circular motions -- we include the full set and adopt the same input data (e.g. distances) in order to be able to compare our analyses with the unconstrained NFW and isothermal fits performed by de\,Blok et al.
The names of the galaxies, their assumed distances, NED types, stellar masses, HI masses (multiplied by 1.39 to correct for He and heavy elements), fitted exponential radial scalelengths for radii where such are meaningful, and the optical diameters (B-band $R_{25}$ from the RC3 catalogue; de\,Vaucouleurs et al. \cite{1991trcb.book.....D}), converted into kiloparsecs) are given in Table\,\ref{table:sample}.
Seven of these galaxies were also contained in the HvAS sample, permitting a direct comparison of the results using better data.

\begin{figure*}
\ifARXIVVERSION
	\includegraphics[width=16cm]{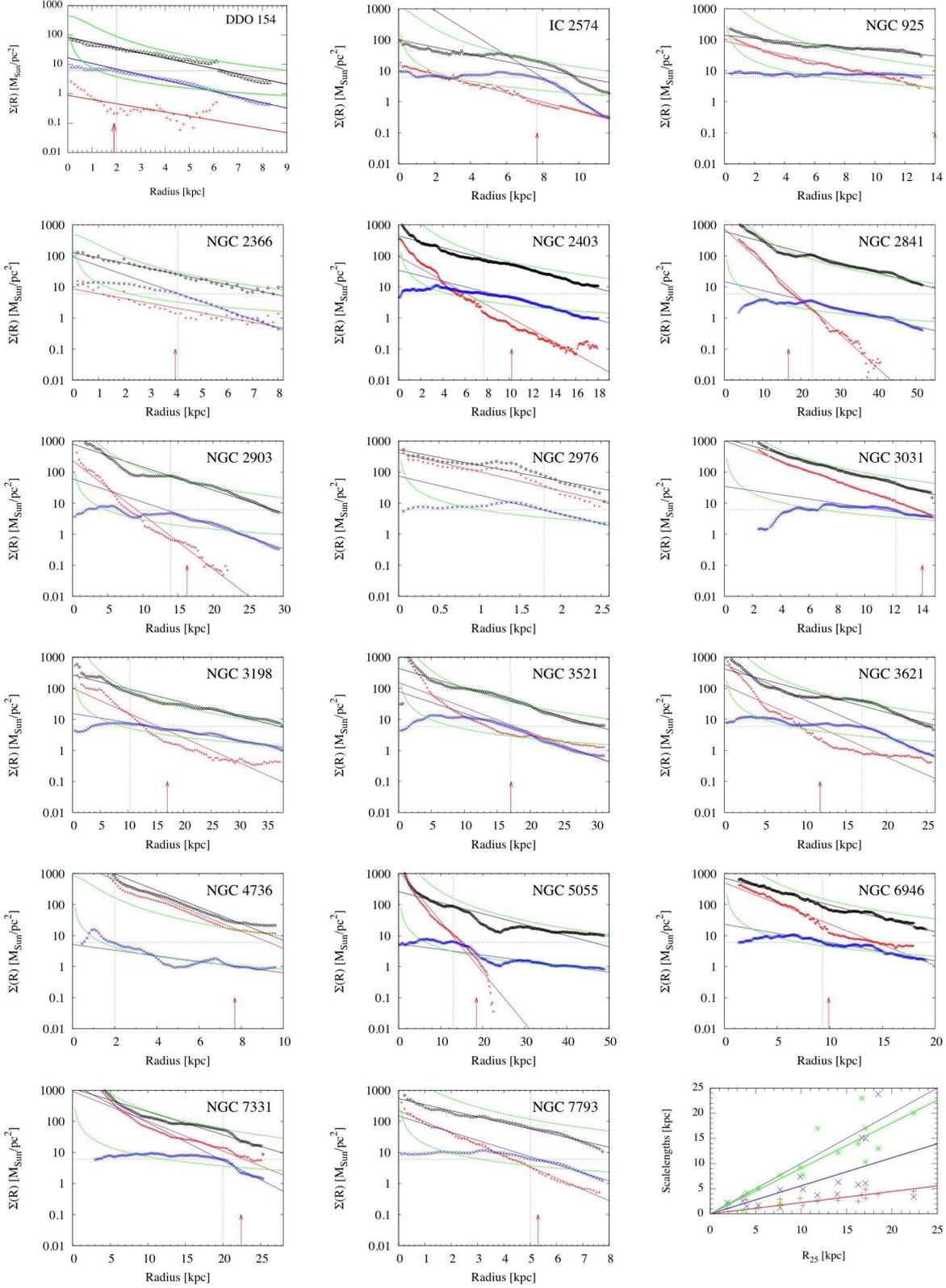}
\else
	\begin{tabular}{cccc}
	\FIGS{5.5}{DDO154_allsigma.eps}{IC2574_allsigma.eps}{NGC925_allsigma.eps} \\
	\FIGS{5.5}{NGC2366_allsigma.eps}{NGC2403_allsigma.eps}{NGC2841_allsigma.eps} \\
	\FIGS{5.5}{NGC2903_allsigma.eps}{NGC2976_allsigma.eps}{NGC3031_allsigma.eps} \\
	\FIGS{5.5}{NGC3198_allsigma.eps}{NGC3521_allsigma.eps}{NGC3621_allsigma.eps} \\
	\FIGS{5.5}{NGC4736_allsigma.eps}{NGC5055_allsigma.eps}{NGC6946_allsigma.eps} \\
	\FIGS{5.5}{NGC7331_allsigma.eps}{NGC7793_allsigma.eps}{R25_scalelengths.eps}
	\end{tabular}
\fi
\caption{Surface densities \& exponential fits (straight lines): 
HI+He (blue exes); stellar disc (red pluses);
implied disc DM (black diamonds); 
$R^{-1}$ power-law fits to the HI and projected URC profiles (green curves);
optical radii $R_{25}$ (red arrows);
radial and surface density range for HI exponential scale lengths $h_{HI}$ (horizontal and vertial light gray dotted lines).  Lower right: comparison of disc size with various scalelengths: HI+He (blue exes), stellar disc (red pluses), inner HI cutoff radii (green asterixes).
\ifARXIVVERSION
	Note that the A\&A version of this figure has the full resolution.
\fi
}
\label{figure:sigmas}
\end{figure*}

The stellar contributions for the THINGS rotation curves were calculated in de\,Blok et al. (\cite{2008AJ....136.2648D}) using $3.6\,\mu m$ images from the {\em Spitzer Space Telescope}, thereby reducing the dependence on ill-constrained extinction corrections.
The mass-to-light ratios, $\Upsilon_{\star}$, were calculated using either a scaled Salpeter (Bell \& de\,Jong \cite{2001ApJ...550..212B}) or Kroupa (\cite{2001MNRAS.322..231K}) initial mass function (IMF): since the scaled Salpeter IMF often resulted in obvious overestimates of the disc contribution in the de\,Blok et al. models, we adopt their Kroupa-based mass-to-light ratios as being more representative of the true stellar components.
The interested reader is directed to the detailed discussions in de\,Blok et al. of the handling of colour gradients and bulge components when calculating the stellar contributions to the rotation curves of individual galaxies.
The resulting mass surface density profiles are shown in Fig.\,\ref{figure:sigmas}
as red pluses (when the $3.6\,\mu m$ data don't extend as far as the HI data, exponential extrapolations are shown).
The exponential scalelengths of the final disc mass-profiles (as opposed to the surface brightness profiles, which are described in de\,Blok et al.) are listed in Table\,\ref{table:sample}.
The tabulated stellar rotation curves generously provided by the THINGS consortium already include the effects of the assumed stellar mass-to-light ratios, i.e. are equal to $\Upsilon_{\star}^{1/2} V_{\star}(R)$, so scaling these velocities immediately yields the Bosma factor $f_{disc}$.

The $\Sigma_{HI+He}$ profiles from the THINGS sample (Fig.\,\ref{figure:sigmas}
blue exes) along with the stellar profiles (red pluses) -- are, in general, not exponential, but can be characterized as roughly flat above a level of $5\!-\!10\,\MSUN\PC^{-2}$, so we have fit an exponential tail to the outer regions either below a level of $6\,\MSUN\PC^{-2}$ (horizontal gray lines in Fig.\,\ref{figure:sigmas})
or outside of an obvious central dip at lower surface densities ; the corresponding scalelengths are given in Table\,\ref{table:sample} and displayed in Fig.\,\ref{figure:sigmas}.
The ratio of this scalelength and the radius of the HI disc varies between 2.0 and 9.8, with a mean value of 4.4, so these HI discs extend farther in terms of exponential scalelengths, than those studied by HvAS.
The correlation between the optical radii, $R_{25}$, the HI scalelengths, and the exponential tail radii, $R_{exp}$, are shown in the bottom right panel of Fig.\,\ref{figure:sigmas}
$R_{exp}$ and $h_{HI}$ are typically $90\pm 7$ and $58\pm 10$ percent of the optical radii, respectively.  
Clearly, the break in the HI distribution occurs exactly at the optical radius, i.e. where gas has been turned into stars.

Also shown in Fig.\,\ref{figure:sigmas}
are $R^{-1}$ power-law fits to the HI data: the only case where this behavior, correponding to the projected density of an isothermal CDM halo, is seen, is perhaps in NGC4736.

The centripetal contributions of the HI gas provided by the THINGS consortium were recalculated from the tabulated total surface mass densities $\Sigma_{HI+He}$ without assuming any extrapolation beyond the last observed radius using the method of Pierens \& H\'ure (\cite{1990AJ....100.1468P}).
While this may lead to a slight overestimation at large radii (Casertano \cite{1983MNRAS.203..735C}), it is impossible to be confident about how to extrapolate the surface densities even given the attempts at defining the exponential tails in Fig.\,\ref{figure:sigmas},
so this treatment makes the least assumptions about the amount and distribution of HI beyond the detection limits.

We note, as has de\,Blok et al. (\cite{2008AJ....136.2648D}; see also Gentile, Famaey \& de\,Blok \cite{2011A&A...527A..76G}), that the rotation curve error bars do not reflect the true statistical errors: they are the result of the non-axisymmetries, spiral arms and patchiness of the HI gas distribution as well as the difficulties of fitting the HI kinematics with a simple model of tilted annulur rings.
Thus, the only comparison we can make is to compare our results externally with previous analyses and internally by comparing the performance of the Bosma effect fits with those assuming CDM.

\begin{figure*}
\ifARXIVVERSION
	\includegraphics[width=16cm]{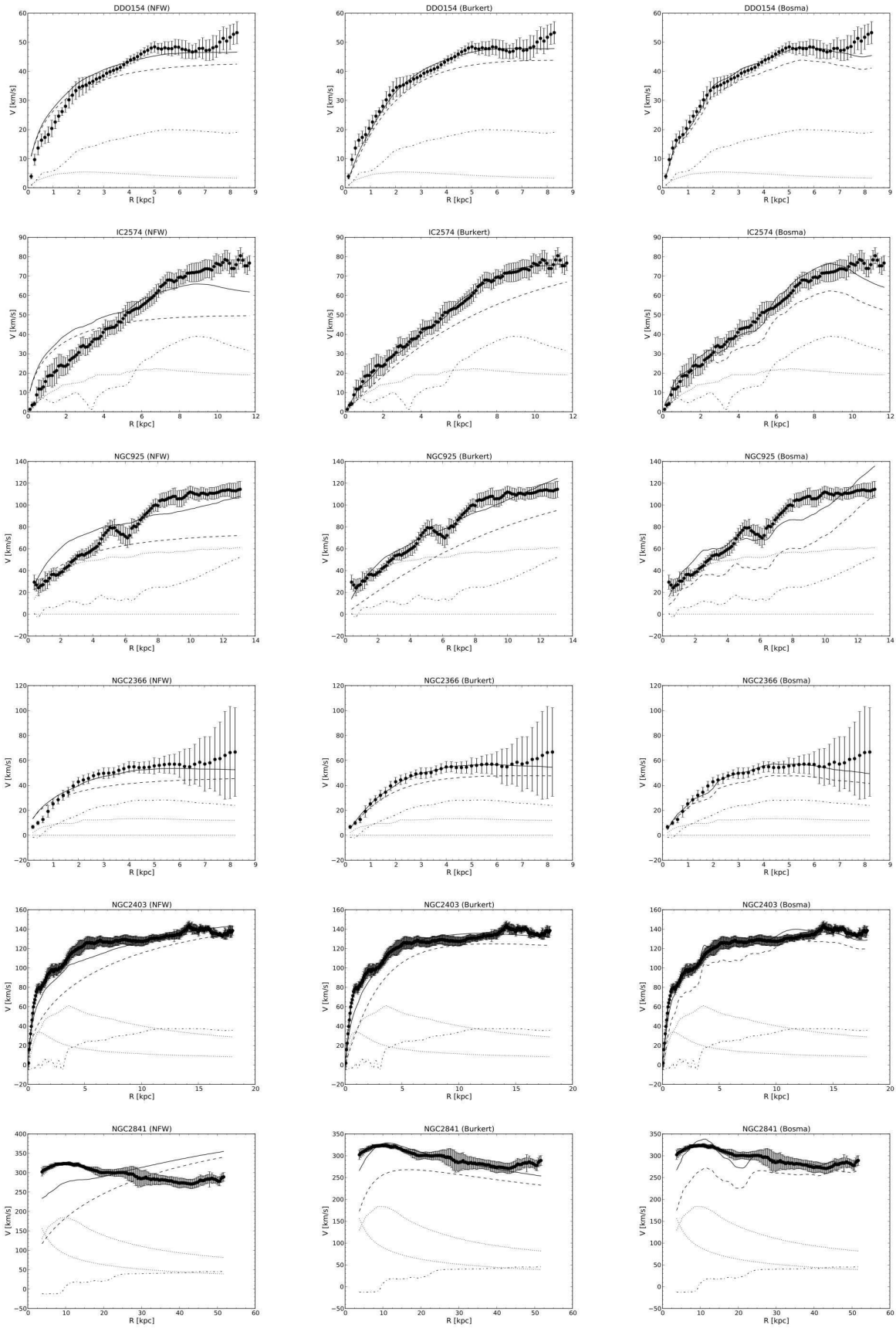}
\else
	\begin{tabular}{cccc}
	\THINGS{DDO154}{_vel_constrainedNFW}{_vel_Burkert}{_vel_Bosma} \\
	\THINGS{IC2574}{_vel_constrainedNFW}{_vel_Burkert}{_vel_Bosma} \\
	\THINGS{NGC925}{_vel_constrainedNFW}{_vel_Burkert}{_vel_Bosma} \\
	\THINGS{NGC2366}{_vel_constrainedNFW}{_vel_Burkert}{_vel_Bosma} \\
	\THINGS{NGC2403}{_vel_constrainedNFW}{_vel_Burkert}{_vel_Bosma} \\
	\THINGS{NGC2841}{_vel_constrainedNFW}{_vel_Burkert}{_vel_Bosma}
	\end{tabular}
\fi
\caption{
Constrained NFW (left), Burkert (centre), and ``classical'' Bosma (right) rotation curve fits.  
The disc and bulge components are plotted as dotted lines, the HI and He contributions as dash-dotted lines, the CDM or Bosma contributions as dashed lines, and the total fitted curve as solid lines.
\ifARXIVVERSION
	Note that the A\&A version of this figure has the full resolution.
\fi
}
\label{figure:v1}
\end{figure*}
\begin{figure*}
\ifARXIVVERSION
	\includegraphics[width=16cm]{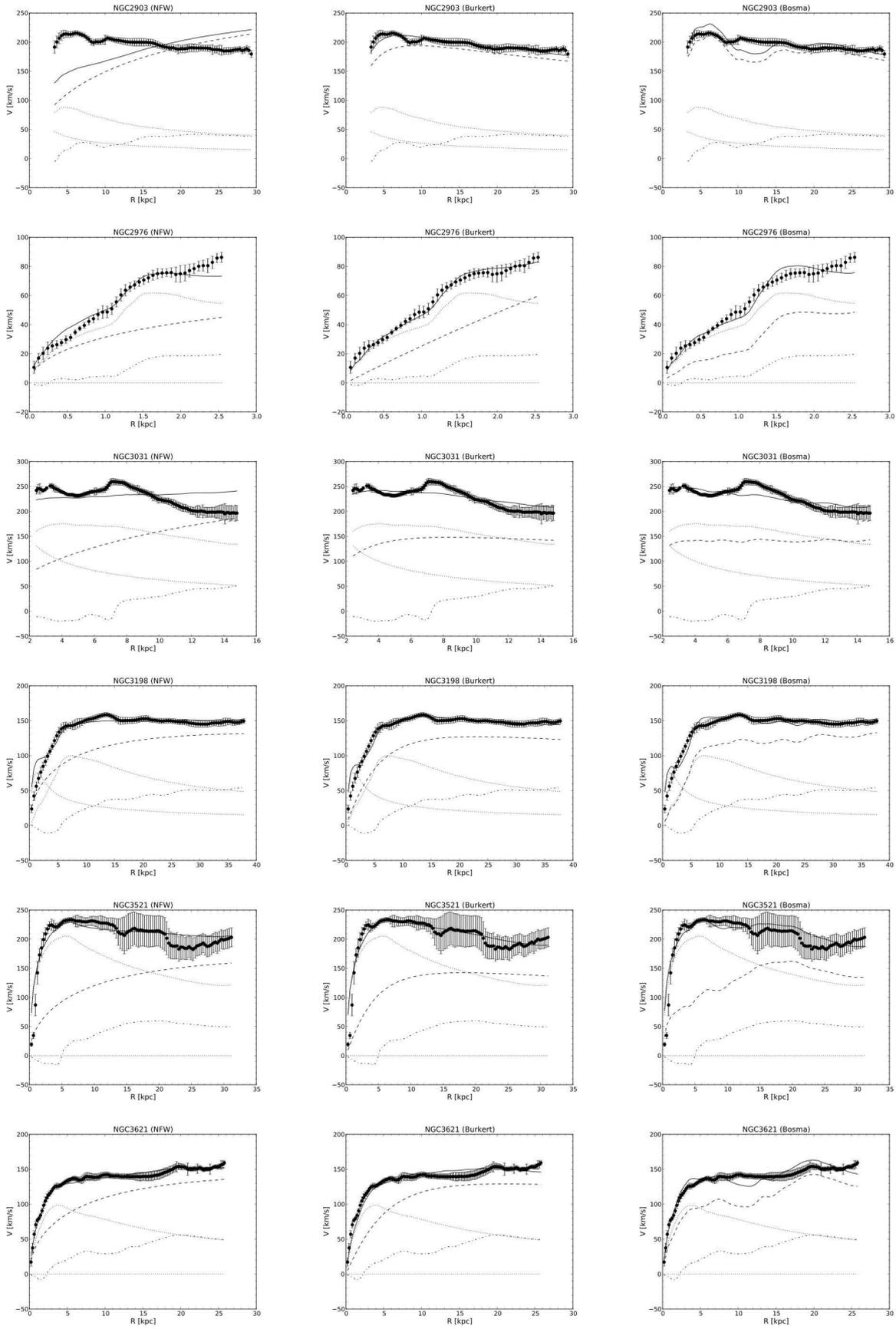}
\else
	\begin{tabular}{cccc}
	\THINGS{NGC2903}{_vel_constrainedNFW}{_vel_Burkert}{_vel_Bosma} \\
	\THINGS{NGC2976}{_vel_constrainedNFW}{_vel_Burkert}{_vel_Bosma} \\
	\THINGS{NGC3031}{_vel_constrainedNFW}{_vel_Burkert}{_vel_Bosma} \\
	\THINGS{NGC3198}{_vel_constrainedNFW}{_vel_Burkert}{_vel_Bosma} \\
	\THINGS{NGC3521}{_vel_constrainedNFW}{_vel_Burkert}{_vel_Bosma} \\
	\THINGS{NGC3621}{_vel_constrainedNFW}{_vel_Burkert}{_vel_Bosma}
	\end{tabular}
\fi
\caption{See caption of Fig.\,\ref{figure:v1}.
\ifARXIVVERSION
	Note that the A\&A version of this figure has the full resolution.
\fi
}
\label{figure:v2}
\end{figure*}
\begin{figure*}
\ifARXIVVERSION
	\includegraphics[width=16cm]{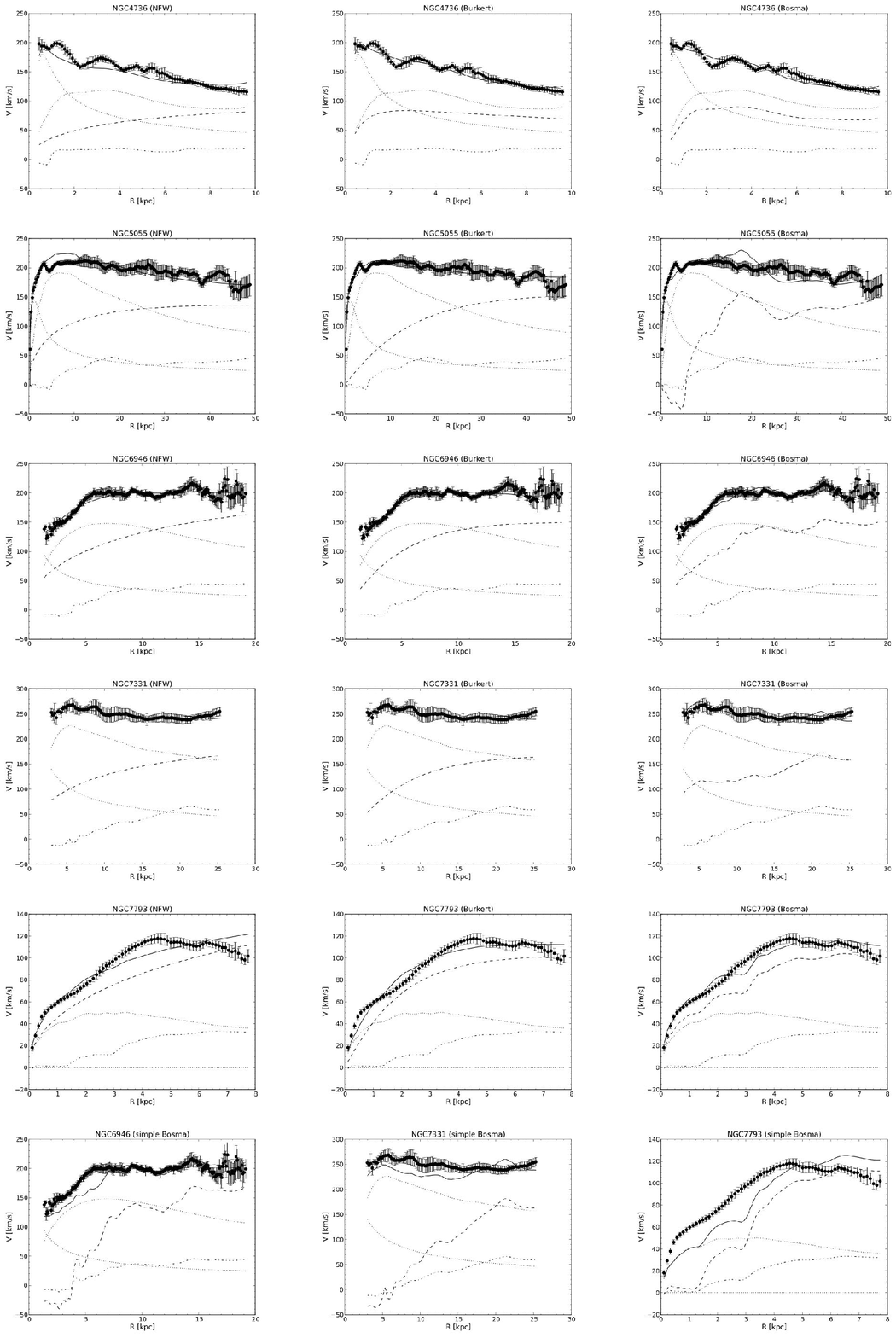}
\else
	\begin{tabular}{cccc}
	\THINGS{NGC4736}{_vel_constrainedNFW}{_vel_Burkert}{_vel_Bosma} \\
	\THINGS{NGC5055}{_vel_constrainedNFW}{_vel_Burkert}{_vel_Bosma} \\
	\THINGS{NGC6946}{_vel_constrainedNFW}{_vel_Burkert}{_vel_Bosma} \\
	\THINGS{NGC7331}{_vel_constrainedNFW}{_vel_Burkert}{_vel_Bosma} \\
	\THINGS{NGC7793}{_vel_constrainedNFW}{_vel_Burkert}{_vel_Bosma} \\
	\FIGS{5.2}{NGC6946_vel_simpleBosma}{NGC7331_vel_simpleBosma}{NGC7793_vel_simpleBosma}
	\end{tabular}
\fi
\caption{See caption of Fig.\,\ref{figure:v1}.
The bottom line shows three ``simple'' Bosma fits for comparison.
\ifARXIVVERSION
	Note that the A\&A version of this figure has the full resolution.
\fi
}
\label{figure:v3}
\end{figure*}


\section{CDM fits}

We first analysed the THINGS rotation curves by performing the constrained NFW and standard URC fits.
The resulting parameters and $\chi^2$ values are shown in Table\,\ref{table:cdm} and the rotation curves in Figs.\,\ref{figure:v1}, \ref{figure:v2} \& \ref{figure:v3}.
In addition, we have listed the virial halo masses, $M_{200}$, the fraction of the CDM halo masses contained within (and hence directly constrained by) the visible galaxy (defined by the last HI rotation curve point), the implied total baryonic mass-fractions, and the core radii.
Detailed descriptions of the CDM fits are given for individual galaxies in Section\,7.

There are clear differences in the median galaxy properties between the two different CDM models.
For instance, the median compactness ratios $c$ are larger (302\%), and the median CDM masses (41\%) and core radii (22\%) smaller for the Burkert/URC models.
Correspondingly, 34\% of the median CDM mass is contained within the region of the visible galaxies for the Burkert/URC models but only 10\% in the constrained NFW models.
Unsurprisingly, the Burkert/URC fits are clearly superior to the constrained NFW fits in almost all cases: the median value of the reduced $\chi^2$ values are 3.6 and 1.2 for the constrained NFW and Burkert/URC fits, respectively.  The most extreme failures of the constrained NFW model occur in DM-dominated galaxies with very flat rotation curves like \object{NGC2841}, \object{NGC2903}, \object{NGC3031}: the correlation between amplitude and shape forces the NFW contributions to have too much radial gradient to maintain the flat shape.
High-mass galaxies like \object{NGC7331} have less problems with constrained NFW profiles because the massive stellar and fitted DM components can be arranged to balance their effects, maintaining the flatness of the rotation curves.
The core radii for the Burkert fits are much smaller than those of the NFW ones, resulting in much smaller CDM masses.

While the CDM models generally result in small baryon-to-CDM mass ratios, the median value for the constrained NFW fits is 4\% versus 10\% for the URC/Burkert fits.  These values are consistent with that of the concordance cosmological model (20.1\%) due to the non-inclusion of molecular, ionized, and infalling gas.
However, if one prefers the URC/Burkert models, one must conclude that these galaxies certainly do not suffer much from the ``missing baryon problem'' (c.f. Bregman \cite{2007ARA&A..45..221B}) and that an apparent dearth of visible baryons can be caused by a systematic overestimate of NFW virial masses.

Interestingly, the CDM fits are not necessarily able to explain the last $\sim$10\% of the rotation curves: even within the unconstrained Burkert/URC fits, there are many cases in which the fits lie significantly either below (\object{DDO154}, \object{NGC2841}, \object{NGC3621}) or above (\object{NGC925}, \object{NGC3031}, \object{NGC7793}) the data.
Part of this problem is undoubtedly due to the difficulty in deriving the mean rotation curves for the often corregated or tilted edges (c.f. Oh et al. \cite{2008AJ....136.2761O}).
This difficulty was one of HvAS criticisms of the Bosma effect -- because they did not test the corresponding performance of CDM models, it was not obvious that this is often a generic problem for any model.


\begin{table*}
\caption{CDM Fits}
\label{table:cdm}
\centering
\begin{tabular}{l | r r r r r r r | r r r r r r r r}
\hline\hline
Name & \multicolumn{7}{|c|}{Constrained NFW (1-parameter)} & \multicolumn{8}{|c}{URC/Burkert (2-parameter)} \\
	 & $V_{200}$ & ${\chi_\nu}^2$ & $c$          & $r_{200}$ & $M_{CDM}$ & $< R_{out}$ & $r_c$
	 & $V_{200}$ & $c$          & ${\chi_\nu}^2$ & $r_{200}$ & $M_{CDM}$ & $< R_{out}$ & $\frac{\displaystyle M_{vis}}{\displaystyle M_{CDM}}$ & $r_c$ \\
\hline
 &&&&&&&&&&&&& \\[-1.8ex]
%
DDO154 	& $34.9\ERR{0.4}$& 3.0	& 10.6	& 50		& 1.4	& 25\%	& 4.7	& $33\ERR{1}$			& $19\ERR{1}$		& 0.6	& 46		& 1.1	& 33\%	& 4\%	& 2.5 \\
IC2574 	& $41\ERR{1}$	& 8.6	& 10.1	& 59		& 2.3	& 29\%	& 5.8	& $71\ERR{4}$			& $10.0\ERR{0.3}$	& 0.2	& 101	& 12		& 10\%	& 7\%	& 10 \\
NGC925 	& $63\ERR{1}$	& 7.1	& 8.9	& 90		& 8.3	& 21\%	& 10		& $112\ERR{11}$			& $10.8\ERR{0.3}$	& 1.2	& 160	& 47		& 7\%	& 2\%	& 15 \\
NGC2366	& $38\ERR{2}$	& 2.3	& 10.4	& 54		& 1.8	& 22\%	& 5.2	& $34\ERR{2}$			& $22\ERR{1}$		& 0.1	& 48		& 1.3	& 34\%	& 8\%	& 2.2 \\
NGC2403	& $139\ERR{1}$	& 2.8	& 7.1	& 198	& 89		& 8.7\%	& 28		& $81.1\ERR{0.4}$		& $28.7\ERR{0.4}$	& 2.0	& 116	& 18		& 36\%	& 4\%	& 4.0 \\
NGC2841	& $394\ERR{3}$	& 43.0	& 5.2	& 563	& 2032	& 6.8\%	& 108	& $154\ERR{1}$			& $42\ERR{1}$		& 2.1	& 219	& 120	& 54\%	& 9\%	& 5.2 \\
NGC2903	& $236\ERR{2}$	& 42.8	& 6.1	& 337	& 437	& 7.2\%	& 56		& $104\ERR{1}$			& $52\ERR{1}$		& 0.8	& 148	& 37		& 52\%	& 5\%	& 2.9 \\
NGC2976	& $54\ERR{3}$	& 2.6	& 9.4	& 76		& 5.1	& 2.3\%	& 8.2	& $136^{+\infty}_{-55}$	& $22\ERR{1}$		& 0.5	& 194	& 83		& 0.3\%	& 0.2\%	& 9.0 \\
NGC3031	& $278\ERR{4}$	& 10.2	& 5.8	& 397	& 714	& 1.7\%	& 69		& $83\ERR{1}$			& $46\ERR{1}$		& 3.5	& 118	& 19		& 37\%	& 33\%	& 2.6 \\
NGC3198	& $117\ERR{1}$	& 2.3	& 7.5	& 167	& 53		& 29\%	& 22		& $94\ERR{1}$			& $18.9\ERR{0.4}$	& 1.3	& 134	& 27		& 49\%	& 16\%	& 7.1 \\
NGC3521	& $150\ERR{4}$	& 5.7	& 6.9	& 215	& 113	& 16\%	& 31		& $97\ERR{3}$			& $25\ERR{1}$		& 4.7	& 138	& 30		& 45\%	& 33\%	& 5.5 \\
NGC3621	& $126\ERR{1}$	& 0.7	& 7.3	& 180	& 67		& 16\%	& 25		& $93\ERR{1}$			& $20.5\ERR{0.3}$	& 2.3	& 133	& 27		& 37\%	& 10\%	& 6.5 \\
NGC4736	& $79\ERR{2}$	& 3.6	& 8.4	& 113	& 16		& 9.1\%	& 14		& $40\ERR{1}$			& $69\ERR{6}$		& 1.4	& 58		& 2.2	& 50\%	& 102\%	& 0.8 \\
NGC5055	& $120\ERR{1}$	& 3.9	& 7.4	& 171	& 57		& 36\%	& 23		& $133\ERR{2}$			& $9.9\ERR{0.2}$		& 1.0	& 190	& 78		& 33\%	& 14\%	& 19 \\
NGC6946	& $182\ERR{2}$	& 1.0	& 6.5	& 260	& 200	& 5.9\%	& 40		& $104\ERR{2}$			& $23.0\ERR{0.5}$	& 1.1	& 149	& 37		& 27\%	& 13\%	& 6.5 \\
NGC7331	& $169\ERR{3}$	& 0.3	& 6.7	& 241	& 160	& 10\%	& 36		& $125\ERR{4}$			& $17\ERR{1}$		& 0.4	& 178	& 65		& 24\%	& 23\%	& 10 \\
NGC7793	& $155\ERR{3}$	& 4.1	& 6.9	& 222	& 125	& 1.8\%	& 32		& $62\ERR{1}$			& $34\ERR{1}$		& 3.1	& 89		& 8.0	& 23\%	& 18\%	& 2.6 \\
\hline
\end{tabular}
\tablefoot{Velocities in $\KMS$, radii in $\KPC$, masses in units of $10^{10}\,\MSUN$, $h=0.7$.}
\end{table*}



\section{The ``simple'' Bosma effect}

\begin{figure*}
\ifARXIVVERSION
	\includegraphics[width=16cm]{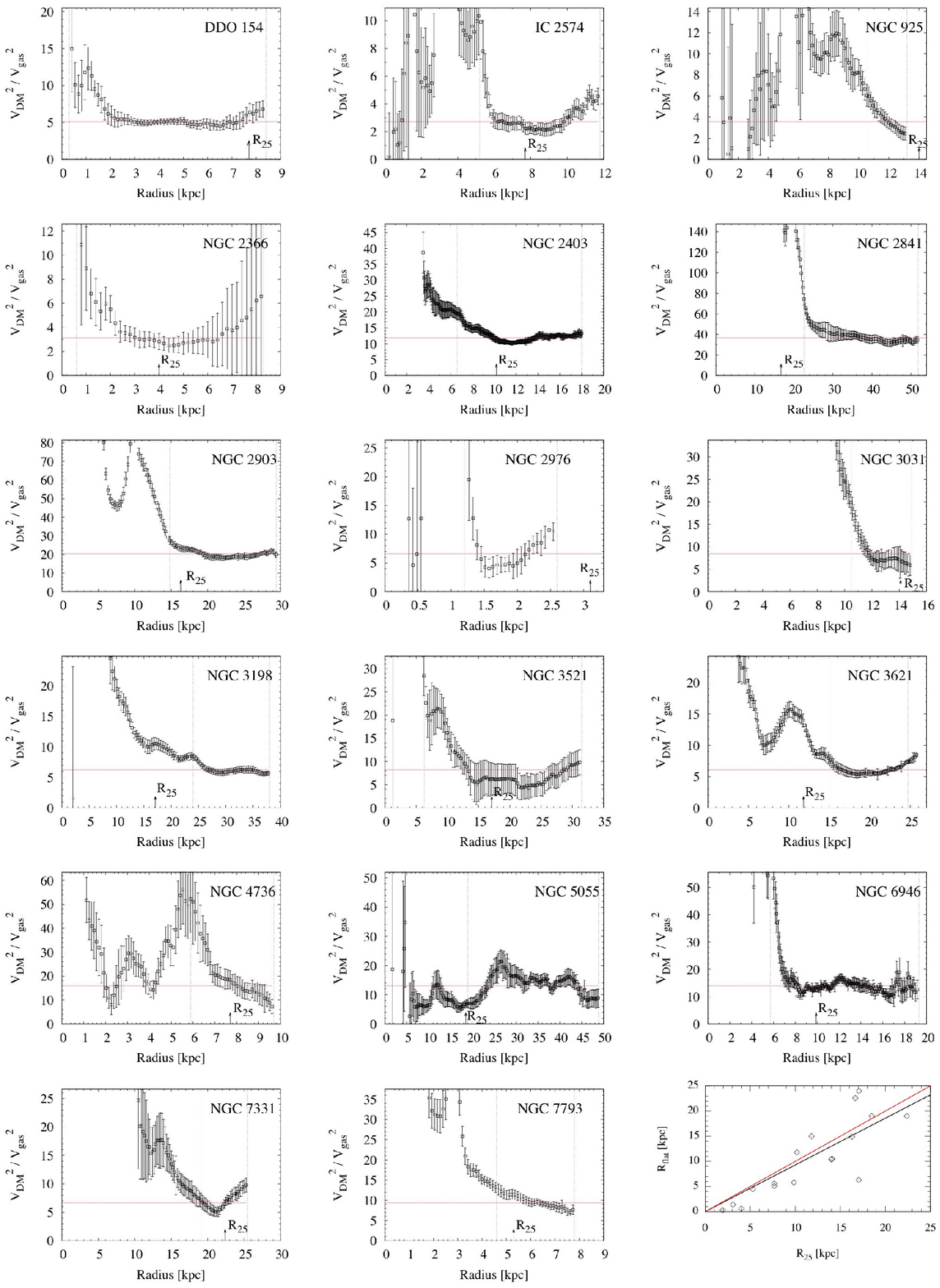}
\else
	\begin{tabular}{cccc}
	\FIGS{5.5}{DDO154_ratio.eps}{IC2574_ratio.eps}{NGC925_ratio.eps} \\
	\FIGS{5.5}{NGC2366_ratio.eps}{NGC2403_ratio.eps}{NGC2841_ratio.eps} \\
	\FIGS{5.5}{NGC2903_ratio.eps}{NGC2976_ratio.eps}{NGC3031_ratio.eps} \\
	\FIGS{5.5}{NGC3198_ratio.eps}{NGC3521_ratio.eps}{NGC3621_ratio.eps} \\
	\FIGS{5.5}{NGC4736_ratio.eps}{NGC5055_ratio.eps}{NGC6946_ratio.eps} \\
	\FIGS{5.5}{NGC7331_ratio.eps}{NGC7793_ratio.eps}{R25_Rflat.eps}
	\end{tabular}
\fi
\caption{Ratio of the centripetal contributions of the DM and HI gas as a function of radius (red horizontal line), corresponding to $1+f_{HI}$ in the ``simple'' Bosma model.  The fitted radial region was selected between the vertical blue lines.
Bottom right: optical radii ($R_{25}$) versus the starting radius where the ratio is statistically flat.
\ifARXIVVERSION
	Note that the A\&A version of this figure has the full resolution.
\fi
}
\label{figure:ratios}
\end{figure*}

We first checked the Bosma effect by plotting the ratio of the centripetal contributions of the DM and gas for each galaxy (the ``simple'' Bosma effect), assuming that the effect is only seen in the HI kinematics:
\begin{equation}
\frac{V^2_{DM}}{V^2_{gas}} \approx \frac{V^2_{obs} - \Upsilon_{\star} V^2_{disc} - \Upsilon_{bulge} V^2_{bulge}}{1.39 V^2_{HI}} \approx 1+f_{HI}(R)
\end{equation}
(see Eqn.\,\ref{equation:model}).
The results, plotted in Fig.\,\ref{figure:ratios},
show that the inner parts of the galaxies have large values of this ratio, i.e. higher rotation than can be explained by scaling the centripetal contribution of HI alone.
This is not surprising:
%
the surface density distributions of HI in spiral galaxies is known to show flat inner regions or even severe dips, resulting in small or even negative contributions to the centripetal acceleration; and
the ``simple'' Bosma model cannot account for the conversion of HI into molecules and stars, which preferentially occurs in the centres of the discs.

What is surprising is the relative constancy of the ratio in the outer discs.
In order to quantify this effect, we fitted the outer ratios with a constant, including just enough data to maintain a reduced $\chi^2 \le 2.0$ (the actual value using realistic errors would be lower).
The radial regions, fractions of the total disc, and resulting Bosma factors are listed in Table\,\ref{table:bosma} (the factors and radial regions are shown as horizontal and vertical lines in Fig.\,\ref{figure:ratios}).
In most cases, the flat regions extend from the optical radius (indicated by vertical arrows in Fig.\,\ref{figure:ratios})
out to the end of the HI disc; a formal fit suggests the flat region starts at $92\pm 8\%$ of the optical radius (the bottom right plot in Fig.\,\ref{figure:ratios}),
i.e. there is a relatively clean ``simple'' Bosma effect signal once one leaves the region dominated by the stars, exactly as originally reported in Bosma \cite{1981AJ.....86.1825B}).
Unsurprisingly, this radius corresponds exactly with $R_{exp}$ (the 8th column in Table\,\ref{table:sample} and the bottom right plot in Fig.\,\ref{figure:sigmas}),
i.e. where the inner HI surface density profile flattens.
This proportionality exists on the average over 54\% of the visible discs, with the value ranging from 19\% (\object{NGC925}) to essentially 100\% (\object{DDO154}).
It is important to note that the relative constancy of the centripetal ratios between DM and HI exists in the outer discs of galaxies independently of the shape of the rotation curve in that region: it applies to flat or slightly downward-sloping rotation curves in massive galaxies as well as to the upward sloping rotation curves of dwarf galaxies.

Given the behavior seen in Fig.\,\ref{figure:ratios},
it is not surprising that the results of formal fits to the entire rotation curves (Table\,\ref{table:bosma}) are poor, since the ``simple'' Bosma fit must be made both for the obviously flat and the obviously non-flat parts of the $V^2_{DM}/V^2_{gas}$ ratio.
This behavior is clearly seen for the three representative fits shown at the bottom of Fig.\,\ref{figure:v3}.
The values of $f_{HI}$ and reduced $\chi^2$ are significantly larger than those of the ``classic'' Bosma models described in the next section.
Thus, ``HI-scaling'' only works in conjunction with a proxy for those regions where the HI surface density is low due to the presence of stars and molecular gas.


\section{The ``classic'' Bosma effect}

The results of ``classic'' Bosma model fits are shown in Figs.\,\ref{figure:v1}, \ref{figure:v2} \& \ref{figure:v3} and in Table\,\ref{table:bosma}.
The quality of the fits, as measured by the reduced $\chi^2$ values relative to those of the URC/Burkert models, is somewhat worse, but it is illuminating to qualify the reasons on a case-by-case basis; this is done for each galaxy in the next section.
In any case, the ``classic'' Bosma fits are better than those of the constrained NFW model, even in those cases where the latter performs adequately.

Comparing our values of $f_{HI}$ with those of HvAS (Table\,\ref{table:bosma}), one sees that similar values are obtained for \object{NGC2403}, \object{NGC2841}, \object{NGC6946}, and \object{NGC7331}.
The origin of the remaining discrepancies is two-fold: first, we have exhaustively probed the available parameter space while HvAS were only interested in finding a solution by hand; and secondly, some of the ``classic'' solutions have values of $f_{disc}$ which are much higher than those which could be normally associated with truly stellar mass-to-light ratios, i.e. extreme ``maximum disc'' solutions.
In our case, this is not an immediate worry, physically, since we are only using the stars as proxies and don't expect the effect to be due to an underestimate of the true stellar mass-to-light ratio.

The values of $f_{disc}$ are generally similar to those seen in ``minimal disc'' fits; the larger values in few galaxies (\object{DDO154}, \object{NGC2366}) are not significant -- the stars play a very subordinate role in these galaxies.
Although the contributions of the stellar and HI proxies are different for each galaxy, the median ratio of the two is 52\%, i.e. the largest contribution is from the ``HI-scaling'' effect.
Thus, it is indeed true that the Bosma effect is primarily a scaling phenomenology between the ISM and DM centripetal contributions.


\begin{table*}
\caption{Bosma Effect Fits}
\label{table:bosma}
\centering
\begin{tabular}{l | r r r | r r | r r r r r r r}
\hline\hline
Name & \multicolumn{3}{|c|}{Outer $V^2_{DM}/V^2_{gas}$ Ratio} & \multicolumn{2}{|c|}{``Simple''} & \multicolumn{7}{|c}{``Classic''} \\
     & $f_{HI}$ & flat over & $R >$                & $f_{HI}$ & ${\chi_\nu}^2$        & $f_{HI}$ & $f_{disc}$ & ${\chi_\nu}^2$ & $f_{HvAS}$ & $M_{dDM}$ & $\frac{\displaystyle M_{vis}}{\displaystyle M_{dDM}}$ & $h_{dDM}$ \\
\hline
 &&&&&&&& \\[-1.8ex]
%
DDO154 	& $4.1\ERR{0.1}$		& 0		& 101\%	& $6.2\ERR{0.1}$		& 1.7	& $5.4\ERR{0.1}$		& $11.2\ERR{1.4}$	& 0.6	& 8		& $0.26\ERR{0.01}$	& 18\%	& 2.4 \\
IC2574 	& $1.7\ERR{0.1}$		& 5.2	& 56\%	& $4.0\ERR{0.1}$		& 4.0	& $2.9\ERR{0.1}$		& $3.3\ERR{0.2}$		& 1.6	& 8		& $2.6\ERR{0.1}$		& 34\%	& 3.7 \\
NGC925 	& $2.6\ERR{0.2}$		& 10.6	& 19\%	& $6.2\ERR{0.1}$		& 4.6	& $4.9\ERR{0.2}$		& $1.40\ERR{0.05}$	& 3.9	& 		& $3.1\ERR{0.1}$		& 37\%	& 9.2 \\
NGC2366	& $2.1\ERR{0.1}$ 	& 0		& 101\%	& $4.1\ERR{0.1}$		& 3.0	& $2.7\ERR{0.3}$		& $6.2\ERR{0.9}$		& 0.4	& 		& $0.33\ERR{0.03}$	& 30\%	& 2.6 \\
NGC2403	& $10.9\ERR{0.1}$	& 6.6	& 63\%	& $13.4\ERR{0.1}$	& 12.2	& $11.0\ERR{0.1}$	& $3.04\ERR{0.04}$	& 1.8	& 10		& $4.96\ERR{0.03}$	& 15\%	& 4.7 \\
NGC2841	& $35.9\ERR{0.6}$	& 22.6	& 56\%	& $50.7\ERR{0.5}$	& 184.4	& $28.0\ERR{0.4}$	& $3.04\ERR{0.02}$	& 2.7	& 23		& $76.2\ERR{0.8}$	& 15\%	& 13.1 \\
NGC2903	& $19.4\ERR{0.3}$	& 14.8	& 50\%	& $28.1\ERR{0.2}$	& 72.5	& $15.2\ERR{0.2}$	& $5.93\ERR{0.08}$	& 4.6	& 37		& $19.5\ERR{0.2}$	& 10\%	& 5.9 \\
NGC2976	& $5.6\ERR{0.5}$		& 1.2	& 51\%	& $8.0\ERR{0.4}$		& 2.7	& $5.2\ERR{0.6}$		& $1.24\ERR{0.04}$	& 1.8	& 		& $0.22\ERR{0.01}$	& 63\%	& 0.9 \\
NGC3031	& $7.4\ERR{0.4}$		& 10.5	& 29\%	& $8.7\ERR{0.3}$		& 75.2	& $4.1\ERR{0.3}$		& $1.69\ERR{0.02}$	& 4.4	& 		& $10.0\ERR{0.1}$	& 62\%	& 3.6 \\
NGC3198	& $5.2\ERR{0.1}$		& 24		& 37\%	& $8.5\ERR{0.1}$		& 19.7	& $6.0\ERR{0.1}$		& $2.13\ERR{0.03}$	& 2.1	& 		& $16.9\ERR{0.2}$	& 25\%	& 10.1 \\
NGC3521	& $7.2\ERR{0.5}$		& 6.3	& 80\%	& $9.7\ERR{0.4}$		& 8.8	& $7.1\ERR{0.4}$		& $1.22\ERR{0.02}$	& 6.1	& 		& $19.8\ERR{0.5}$	& 50\%	& 5.8 \\
NGC3621	& $5.1\ERR{0.1}$		& 15		& 38\%	& $8.3\ERR{0.1}$		& 19.6	& $6.8\ERR{0.1}$		& $1.74\ERR{0.02}$	& 4.2	& 		& $11.2\ERR{0.1}$	& 24\%	& 6.1 \\
NGC4736	& $14.8\ERR{1.0}$	& 5.9	& 40\%	& $19.9\ERR{0.6}$	& 2.9	& $3.0\ERR{1.7}$		& $1.53\ERR{0.05}$	& 1.7	& 		& $3.6\ERR{0.9}$		& 61\%	& 2.0 \\
NGC5055	& $12.0\ERR{0.3}$	& 19		& 61\%	& $12.5\ERR{0.2}$	& 3.1	& $12.9\ERR{0.2}$	& $0.97\ERR{0.02}$	& 3.1	& 		& $26.4\ERR{0.3}$	& 41\%	& 11.6 \\
NGC6946	& $12.8\ERR{0.2}$	& 5.7	& 71\%	& $15.1\ERR{0.1}$	& 5.5	& $9.9\ERR{0.2}$		& $1.40\ERR{0.02}$	& 1.3	& 5		& $11.2\ERR{0.1}$	& 45\%	& 4.9 \\
NGC7331	& $5.7\ERR{0.2}$		& 19		& 24\%	& $8.5\ERR{0.2}$		& 4.0	& $6.2\ERR{0.2}$		& $1.27\ERR{0.02}$	& 1.0	& 8		& $24.8\ERR{0.4}$	& 59\%	& 5.9 \\
NGC7793	& $8.4\ERR{0.3}$		& 4.6	& 42\%	& $12.8\ERR{0.2}$	& 26.5	& $8.9\ERR{0.2}$		& $2.42\ERR{0.04}$	& 2.4	& 		& $11.1\ERR{0.2}$	& 13\%	& 2.2 \\
\hline
\end{tabular}
\tablefoot{Radii in $\KPC$, masses in units of $10^{10}\,\MSUN$, $f_{HvAS}$ are $f_{HI}$ from HvAS.}
\end{table*}


\section{Results for individual galaxies}

Here is a detailed description of the CDM and Bosma effect fits to individual galaxies.\\

{\it \object{DDO154}} has a rotation curve typical of dwarf galaxies, with a rapid rise in the inner disc but no asymptotic flattness.
The observed rotation curve is flatter than the constrained NFW fit for radii $>6\,\KPC$ and higher than either the Burkert/URC or ``classic'' Bosma fit for the last $\sim$10\%.
Thus, no model is able to fit the exact behavior of the outer disc.
The ``classic'' Bosma fit yields a very accurate fit to most of the rotation curve, including the ``dents'' in the inner curve which occur at radii of $\sim$1/2, 2, and $5\,\KPC$.  The large nominal value of $f_{disc}$ is not significant -- the stars play practically no role in the fit.

The rotation curve of {\it \object{IC2574}} rises continuously over the entire disc, like that of \object{DDO154}.
The constrained NFW fit overpredicts the rotation curve at small radii.
The mass of the Burkert/URC cannot be determined due to the severe degeneracy of the model parameters.
Both the constrained NFW and ``classic'' Bosma fits fall short of the rotation curve at the outer edges.
Because of the non-smooth distribution of HI, the Bosma fits have too much small-scale structure, resulting in a significant difference in the value of $\chi^2$ relative to the Burkert models.  Nevertheless, the Bosma fit is still quite good (${\chi_\nu}^2=1.6$).

The distribution of HI in {\it \object{NGC925}} is totally flat and definitely cannot be approximated by an exponential.
This rotation curve rises continuously at large scales but shows prominent ``bumps'' at $5-6\,\KPC$.
The latter cannot be explained by any model.
Again, the mass of the Burkert/URC cannot be determined due to the severe degeneracy of the model parameters.
Both the constrained NFW and the ``classic'' Bosma fits overestimate the curve at small radii and underestimate it at larger radii.
Correspondingly, the Bosma model overpredicts the rotation curve at the disc edge; this is exactly the opposite effect than that expected by HvAs.

{\it \object{NGC2366}} has a classic rotation curve that becomes asymptotically flat (though the errors are large for the outer disc).
All of the models can reproduce the data equally well.  As in \object{DDO154}, the stellar part of the ``classic'' fit plays no role, so the large value of $f_{disc}$ is not significant.

The rotation curve in {\it \object{NGC2403}} rises rapidly and becomes flat/slowing rising for about 3/4 of the visible disc.
The constrained NFW fit systematically underpredicts the inner and overpredicts the outer halves.
The Burkert/URC fit is much better but underpredicts in innermost disc.
The ``classic'' Bosma fit does much better in the inner third of the disc, following more of the deviations from a simple, smooth curve.
None of the models can follow the ``bumps'' in the outer half of the disc.

The rotation curve of {\it \object{NGC2841}} has a maximum around $10\,\KPC$ and very slowly declines out to the edge.
Again, the constrained NFW fit produces a systematically false slope.
The Burkert/URC fit does much better, though at the price of a very small core radius (10\% of the galaxy radius and 1/3 of the HI scalelength!) and it also cannot explain the increase in the last 10\% of the disc.
The ``classic'' Bosma model does a much better job in the outer disc, roughly follows the shape but shows too much structure in the inner 1/2 of the disc, resulting in a slightly higher value of ${\chi_\nu}^2 = 2.7$ compared with the URC fit ($2.0$).

The results for {\it \object{NGC2903}} closely follow those of \object{NGC2841}: the smaller scale wiggles in the ``classic'' Bosma fit produce the big difference in ${\chi_\nu}^2$ (4.6 versus 0.8) even though Bosma model reproduces the basic flat shape fairly well.

{\it \object{NGC2976}} is another dwarf galaxy with a rotation curve similar to \object{IC2574}.
The mass of the Burkert/URC halo cannot be determined due to the degeneracy of the model parameters: NGC2976 need a roughly linearly increasing centripetal contribution that can be provided by practically any large value of $V_{200}$.
All models do a good job of reproducing the rotation curve, whereby the Burkert/URC is nominally the best and, again, the ``classic'' Bosma fit shows too much small-scale structure.

{\it \object{NGC3031}} (\object{M81}) has a rotation curve very similar to that of \object{NGC2841}, showing a fairly flat shape accompanied by a globally declining level and a big ``bump'' at $\sim 7\,\KPC$ due to the prominent spiral arms.
The constrained NFW model predicts an extremely flat rotation curve, whereas both the Burkert/URC and the ``classic'' Bosma models can explain the gentle drop to larger radii but cannot explain the ``bump''.

{\it \object{NGC3198}} has a classic rotation curve that rapidly increases to a very flat level. 
The amplitude of the constrained NFW fit doesn't permit a small enough core radius to flatten the CDM contribution.
All of the models have a problem with the bulge, but the ``classic'' Bosma model attempts to compensate by using the negative centripetal effect of the HI contribution, resulting in an only slightly worse fit that that of the Burkert/URC model.

{\it \object{NGC3521}} is very similar to \object{NGC2841} and \object{NGC2903}.
The constrained NFW model again cannot produce a flat enough contribution.
Both the Burkert/URC and the ``classic'' Bosma models can fit the outer disc but have a difficult time in the inner 10\% of the disc, resulting in ${\chi_\nu}^2$ values of 4.7 and 6.1, respectively.

{\it \object{NGC3621}} is very similar to \object{NGC3521}.
The ``classic'' Bosma fit does a better job in the inner disc but shows too much smaller-scale structure in the outer disc, resulting in a higher value of ${\chi_\nu}^2$ (4.2 versus 2.2 for Burkert/URC).
Both the Burkert/URC and the Bosma models underpredict the outer rotation curve.

The rotation curve of {\it \object{NGC4736}} is very similar to that of \object{NGC3031} but has more ``bumps'' with smaller amplitude.
Again, the constrained NFW fit is systematically low in the inner and high in the outer disc.
The Burkert/URC and ``classic'' Bosma fits are essentially equally good, but the former needs a very small core radius.

{\it \object{NGC5055}} is very similar to \object{NGC3521}: while the CDM models do an adequate job, the Bosma models suffer from the smaller-scale variations in the HI centripetal contribution due to the highly structured HI disc.

The strong stellar component in {\it \object{NGC6946}} permits even the constrained NFW model to do a reasonable job at fitting this rotation curve -- all fits have roughly the same value of ${\chi_\nu}^2$ (1.4, 1.1, and 1.3 for the constrained NFW, Burkert, and ``classic'' Bosma models, respectively).
Nevertheless, the ``kink'' at $6\,\KPC$ separating the inner rise with the outer flat curve is best reproduced by the Burkert/URC and ``classic'' Bosma models.
The latter is the best at explaining the other ``bumps'' in the outer half of the disc.

{\it \object{NGC7331}} is very similar to \object{NGC2841}, \object{NGC2903}, and \object{NGC4736}.
All of the models do a very good job of explaining the rotation curve.
Only the constrained NFW model is increasing enough at the outer radius to explain the observed slight increase at the edge.

{\it \object{NGC7793}} is similar to \object{DDO154}, although the rotation curve slopes downward in the outer half of the disc, resulting in a poor constrained NFW fit.
The Burkert/URC model cannot reproduce the ``dent'' at $2\,\KPC$ (or, equivalently, the ``peaks'' at 0.5 and $4.5\,\KPC$).
While the ``classic'' Bosma model shows too much small-scale structure, its ability to follow the medium-scale structure results in the best value of ${\chi_\nu}^2$ (2.4 versus 5.5 and 3.1 for the CDM models).\\

In summary, the generic Burkert/URC model is most successful -- as expected.
The constrained NFW model is rarely as successful as the Burkert/URC or ``classic'' Bosma models, a situation explained by the ``cuspiness'' of the basic profile as well as the strict correlation between the amplitude and the compactness.
The ``classic'' Bosma model does remarkably well: in most cases, the main source of failure of the latter is the presence of too much small-scale structure created by the mottled appearance of the HI disc (spiral arms, etc.).
In fact, one has the impression that an appropriate smoothing of the centripetal contributions at small scales would result in nearly perfect rotation curves in practically all cases.


\section{Discussion}

We can now review the arguments presented by HvAS against the reality of the ``classic'' Bosma effect.
\begin{itemize}
\item ``{\it The model curve} [of the poorer fits] {\it does not agree with the observed rotation curve in the inner region.}''
Our results show that the ``classic'' Bosma model does not have this problem any more than the CDM models.
On the contrary, the Bosma model is often much more successful in this region.  
Given the fact that we have used the stars as proxies, this is perhaps not surprising: we have in effect used the known efficacy of ``maximal disc'' models to compensate for the negative centripetal contributions of the HI discs, the latter produced by the non-exponential HI surface density distributions.
\item There are ``{\it ... large wiggles that are not present in the observed rotation curve.}''
This effect, when present, is easily explained as being due to the inadequacy of using HI as a total mass tracer at small spatial scales.  On the other hand, there are galaxies where the observed ``wiggles'' in the rotation curve are more easily explained by the Bosma model, where $V(R)$ is more closely tied to the gas density, rather than by a CDM model whose small-scale density variations must be small (e.g. Springel et al. \cite{2008MNRAS.391.1685S}).
\item ``{\it The model rotation curve drops below the observed rotation curve at large radii.}''
This effect is obviously present relative to the performance of the best CDM model in only 2 of the 17 rotation curves (IC2574, NGC2976); the other weaker examples show variations in both directions of a magnitude corresponding to systematic variations from a smooth rotation curve and hence not specific to the Bosma model (there are also cases where the model lies above the rotation curve).  The difference between the CDM and Bosma models, while present, is not large.  If the HI suffers from ionisation and/or formation of H$_2$ in the outer disc, this minor effect could easily be due only to the ISM chemistry and dynamics in the outer, relatively starless disc.
\item ``{\it ... scaling of HI to represent the dark component only works in combination with maximal discs.}''
The flatness of or dips in the central HI surface mass-density distributions creates centripetal deficits which force the use of a maximal disc when fitting ``classic'' Bosma models.  This is obviously due to the inadequacy of using the neutral component of the ISM alone and is thus expected: {\it the relative success of maximal-disc models is a natural consequence of the ``classical'' Bosma effect and not a sign of its failure}.
\item ``{\it ... our sample is biased against galaxies with $R_{out}/h_{HI}$ substantially larger than 3.}''  This argument does not apply to the new data sets used: the mean value of $R_{out}/h_{HI}$ for the THINGS sample is 4.4, with values between 2 and 10 present.  In any case, this argument by HvAS was originally based on the assumption, that the HI is distributed exponentially, whereas the observed distribution looks more like an outer exponential with an inner disc saturated at surface densities around $5-10\,\MSUN\,\PC^{-2}$.
\item While HvAS found that ``{\it .. for about two-thirds of the galaxies} [they] {\it obtain good fits to the data}'', we find a much higher success rate when comparing with the corresponding CDM models.
Formally, the fraction of fits having values of the reduced $\chi^2$ less than 2 is 18\% for the NFW, 70\% for the Burkert/URC, and 47\% for the ``classic'' Bosma models.  The median values of ${\chi_\nu}^2$ are 3.6, 1.6, and 2.6 for the constrained NFW, Burkert/URC, and ``classic'' Bosma models, respectively.
That the nearly unconstrained URC models perform better than the highly constrained Bosma models is, after all, not surprising.
Given that the systematic errors in the observations which produce the sometimes highly structured rotation curves are not included in the values of $\chi^2$, one can conclude that the Bosma model is, in fact, quite adequate.
Thus,
\item we find that it is not at all true that ``{\it the good fits are somewhat coincidental}''.
\end{itemize}

We conclude that HvAS conservative rejection of the Bosma effect was premature and based upon a too subjective judgement of the data.
While the rotation curve fits are not perfect, they hold up very well even against CDM fits with considerable freedom in the amplitude and shape of the DM components -- and this despite the severe inflexibility of shape present in the ``classic'' Bosma models.
Wherever the effect appears to fail, there are either usually good reasons why a trivial interpretation of the failure is misleading or signs that the problem does not occur for the Bosma effect alone, but is a problem for any mass model.


\section{The cause of the Bosma effect?}

Given that the Bosma effect appears to be real, the obvious question is why it occurs.
The standard explanation would be that this is yet another correlation between the effects of baryons and the distribution of CDM.
However, the explicit distinction between the old (Eqn.\,\ref{equation:fHI}) and new definitions of the Bosma ratio (Eqn.\,\ref{equation:fISM}) is not at all subtle: whereas it was originally plausible to think of an unknown DM component imbedded within and hence proportional to the baryonic disc, leading to the definition in Eqn.\,\ref{equation:fHI}, the simultaneous placement of the DM in a quasi-spherical halo and the requirement of direct proportionality of its gravitational effects with a dynamically minor component in a disc constitutes a difficult problem in an individual galaxy and a seemingly insurmountable problem for a large and dynamically heterogeneous sample of galaxies.
We have seen that the supposed connection between the projected CDM densities and the HI densities mentioned in the literature is a myth based on a misinterpretation of the observational basis for the Bosma effect.
The observed correlation is thus either a tracer of a totally inexplicable interaction between CDM and the baryonic disc or a telling argument against the paradigm of CDM in a non-disc halo.

If one interprects the Bosma effect literally, as being due to the presence of disc DM in some form associated with the ISM, the amount of additional mass is significant but not totally unreasonable.
The total implied disc DM masses, $M_{dDM}$, and the corresponding visible-to-disc DM ratios 
\begin{equation}
\frac{M_{vis}}{M_{dDM}} \equiv \frac{M_{disc} + M_{bulge} + M_{HI+He}}{f_{disc} M_{disc} + f_{HI} M_{HI+He}}
\end{equation}
in Table\,\ref{table:bosma} suggest that the visible mass makes up 11-63\% of the total, with a mean of 25\%.
The derived distributions of the Bosma component are also shown in Fig.\,\ref{figure:sigmas}
(black diamonds): one has the striking impression that they are exponential with large scalelengths more often related to the outer HI disc than to the stellar disc.
The ``classic'' Bosma effect appears to conspire so that the dent in an exponential gaseous disc visible in the HI is filled up by a scaled version of the stellar disc, revealing a globally exponential distribution of disc DM.
%
Also shown in Fig.\,\ref{figure:sigmas} are the projected surface densities of the Burkert/URC fits: with the exception of NGC2841, the disc DM -- if it exists -- is not distributed like the projected CDM halos.

A candidate for disc DM that would naturally explain the Bosma effect is a nearly invisible ISM component consisting of very dense and cold H$_2$ cloudlets (Pfenniger, Combes \& Martinet \cite{1994AA...285...79P}; Pfenniger \& Combes \cite{1994AA...285...94P}; Gerhard \& Silk \cite{1996ApJ...472...34G}) whose dynamical effects within the galaxies could be very different from those usually assumed for the visible matter (Revaz et al. \cite{2009A&A...501..171R}).
Indeed, there are many reasons based on the formation and dissociation of H$_2$ to believe that the visible HI structures in galaxies belie only a fraction of the true gaseous content, fully independently of any questions concerning their centripetal signatures (Allen \cite{2004ASSL..319..731A}).
On the other hand, it is generally argued that it is impossible for the discs of spiral galaxies to contain enough matter to explain the rotation curves for a variety of reasons, making it easy to dismiss this interpretation of the Bosma effect on largely theoretical grounds.
This is not the adequate place to address each of these arguments in detail -- this will be the subject of a future paper in this series --  but it is clear that they are based upon sets of assumptions or interpretations that may sound reasonable but are not guaranteed to be correct, at least as far as the necessity of exotic {\it cold} DM is concerned (warm or hot DM cannot be adequately probed at the small scales used here: Sanders \cite{2007MNRAS.380..331S}; Gentile, Zhao \& Famaey \cite{ 2008MNRAS.385L..68G}).
In any case, the mass needed to explain the Bosma effect using disc DM is still much lower than the amount of ``missing baryons'' within the concordance $\Lambda$CDM model (c.f. Bregman \cite{2007ARA&A..45..221B}), so the question is not one of having enough baryons but whether one can place them within stable galaxy discs in a form that is not easily observed.

If the Bosma effect is telling us that the baryons have simply not been given enough weight, then the only other alternatives are a change in the properties of gravity like MOND (see Gentile, Famaey \& de\,Blok \cite{2011A&A...527A..76G} for an analysis of the same data) and Conformal Gravity (Mannheim \& O'Brien \cite{2011PhRvL.106l1101M}) or the presence of non-linear effects in self-gravitating systems within General Relativity (Cooperstock \& Tieu \cite{{2005astro.ph..7619C}}; Carrick \& Cooperstock \cite{2011arXiv1101.3224C}; but see Fuchs \& Phleps \cite{2006NewA...11..608F}), possibly resulting in non-Newtonian behavior even in low-gravity systems.
Which of these different explanations one finds objectionable, plausible or appealing is, of course, a matter of perspective.

The Bosma effect is a significant addition to the other known but poorly understood correlations between the baryonic components of spiral galaxies and their DM components.
As such, it deserves much more study.
In following papers, we will make a better correction for the amount of matter in the ISM in order to further constrain the origin of this effect
and present a comparison of the Bosma effect with other baryon-DM correlations along with a critical review of the classic arguments against disc DM.


\begin{acknowledgements}
This work made use of ``The HI Nearby Galaxy Survey'' (Walter et al. 2008).  We would like to thank E. de\,Blok and the rest of the THINGS consortium for generously providing us with their extracted rotation curves and surface density distribution data as well.
This paper profitted from discussions with many colleagues, particularly E. de\,Blok and P. Kroupa.
This research has made use of the NASA/IPAC Extragalactic Database (NED), which is operated by the Jet Propulsion Laboratory, California Institute of Technology, under contract with the National Aeronautics and Space Administration, and the {\em Vizier} catalogue database, which is operated by the Centre de Donn\'ees astronomiques de Strasbourg.
\end{acknowledgements}


\end{document}